\documentclass[fleqn,10pt]{SelfArx} 

\definecolor{color1}{RGB}{0,0,90} 
\definecolor{color2}{RGB}{0,20,20} 

\newlength{\tocsep} 
\usepackage{lipsum} 

\usepackage{mathptmx} 
\usepackage{comment} 
\usepackage[T1]{fontenc}

\usepackage[affil-it]{authblk}

\usepackage{amsmath}
\usepackage{amssymb}
\usepackage{graphicx}
\usepackage{subfigure}
\usepackage{color}

\JournalInfo{Scientific Reports} 
\Archive{doi:10.1038/srep01645} 

\PaperTitle{The predictability of consumer visitation patterns} 

\Authors{Coco Krumme\textsuperscript{1}, Alejandro Llorente\textsuperscript{3}*, Manuel Cebrián\textsuperscript{2,6}, Alex ("Sandy") Pentland\textsuperscript{1}, Esteban Moro \textsuperscript{3,4,5}}

\affiliation{\textsuperscript{1}\textit{Media Laboratory, Massachusetts Institute of Technology, Cambridge, MA 02139, USA}} 
\affiliation{\textsuperscript{2}\textit{Department of Computer Science and Engineering, University of California at San Diego, La Jolla, CA 92093, USA}} 
\affiliation{\textsuperscript{3}\textit{Instituto de Ingenier\'{\i}a del Conocimiento, Universidad
  Aut\'onoma de Madrid, Madrid 28049, Spain}} 
\affiliation{\textsuperscript{4}\textit{Departamento de Matem\'aticas \& GISC, Universidad Carlos
  III de Madrid, Legan\'es 28911, Spain}} 
\affiliation{\textsuperscript{5}\textit{Instituto de Ciencias Matem\'aticas
  CSIC-UAM-UCM-UC3M, Madrid 28049, Spain}} 
\affiliation{\textsuperscript{6}\textit{NICTA, Melbourne, Victoria 3010, Australia}} 
\affiliation{*\textbf{Corresponding author}: alejandro.llorente@iic.uam.es} 

\Abstract{
We consider hundreds of thousands of individual economic transactions to ask: how predictable are consumers in their merchant visitation patterns? Our results suggest that, in the long-run, much of our seemingly elective activity is actually highly predictable. Notwithstanding a wide range of individual preferences, shoppers share regularities in how they visit merchant locations over time.  Yet while aggregate behavior is largely predictable, the interleaving of shopping events introduces important stochastic elements at short time scales. These short- and long-scale patterns suggest a theoretical upper bound on predictability, and describe the accuracy of a Markov model in predicting a person's next location.  We incorporate population-level transition probabilities in the predictive models, and find that in many cases these improve accuracy.  While our results point to the elusiveness of precise predictions about where a person will go next, they suggest the existence, at large time-scales, of regularities across the population.\\
}

\newcommand{\keywordname}{Keywords} 

\begin{document} 
\maketitle


\section{Introduction}

Human economic behavior is curbed by human geography: constraints on 
mobility determine where we can go and what we can buy. At the same 
time, the electiveness of shopping itself drives our movement.  
Understanding consumer patterns is important not only for modeling 
the dynamics of a market, but also in discovering how past 
performance predicts future behavior at the individual level. We are concerned less with predicting how much is spent or what is bought, and rather with where a person will go next.

Economic models of consumption incorporate constraint and choice 
to varying degrees. In part, shopping is driven by basic needs and constraints, 
with demand shaped by price, information and accessibility \cite{nelson1970information,belk1975situational}.  At the same time, it
 is believed that shoppers will opt for greater variety if possible 
\cite{stiglitz}, although empirical work finds that 
behaviors such as choice aversion \cite{jams} and brand-loyalty can 
limit search
\cite{jacoby1978brand,muniz2001brand,bonfield1974attitude}. How do choice and constraint connect? Investigations
with
mobile phone data find that individual trajectories are largely predictable
\cite{Brockmann2006, gonzalez2008understanding,bagrow2012mesoscopic,lu2012predictability}.
Yet these models say little about the motivation for movement. At the same time, models of small-scale decision-making \cite{ryan1975fishbein,wilson1975empirical,ajzen1977attitude,miniard1983modeling} leave open the question of how individual heuristics might form large-scale patterns.

Here, we draw on a unique set of individual shopping data, with tens of thousands of individual time series representing a set of uniquely identified merchant locations, to examine 
how choice and necessity determine the predictability of human behavior. 
Data from a wealth of sensors might be captured at some arbitrary 
waypoint in an individual's daily trajectory, but a store is a destination, 
and ultimately, a nexus for human social and economic activity.

We use time series of de-identified credit card accounts from two major financial 
institutions, one of them North American and the other European. Each account
 corresponds to a single individual's chronologically ordered time series of purchases over 6-11 months, revealing not only how much money he spends, but how he allocates 
his time across multiple merchants. 
We filter
for time series most likely to represent real individuals
rather than cards used by multiple people, or cards representing
infrequent or corporate usage. This leaves us with time series with at least 10 but no more than 50 unique stores per
month, as well as at least 50 but no more than 120 purchases per
month. The data is further described in the Methods section.

To quantify the predictability of shopping patterns, we compare individuals 
using two measures. First, we consider static predictability, using 
temporally-uncorrelated (TU) entropy to theoretically bound, and a frequentist model to 
predict where a person will be. Second, we consider a person's dynamics, 
by taking into account the sequence in which he visits stores. Here we use an 
estimate of sequence-dependent (SD) entropy to measure, and a set of Markov Chain models to predict location. Both entropy measures, and predictive model, are defined fully below.

\section{Results}

\begin{figure*}
\centering
\includegraphics[width=\textwidth]{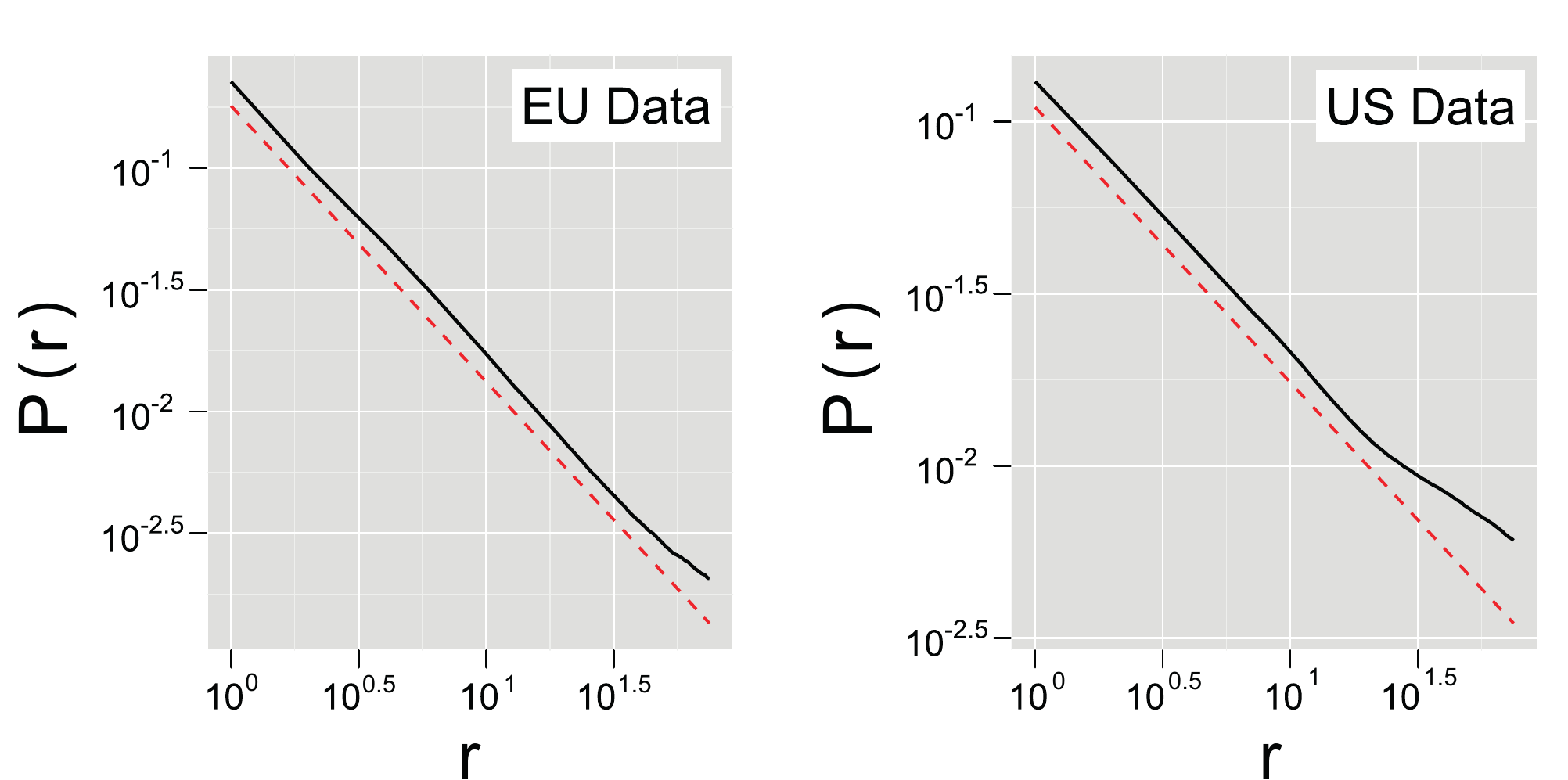}
\caption{Probability of visiting a merchant, as a function of merchant
  visit rank, aggregated across all individuals. Dashed line
  correspond to power law fits $P(r) \sim r^{-\alpha}$ to the initial
  part of the probability distribution with $\alpha = 1.13$ for the
  European and $\alpha = 0.80$ for the North American database. \label{zipf}}
\end{figure*}

At longer time scales, shopping behavior is constrained by some of the same 
features that have been seen to govern human mobility patterns.
We find that despite varied individual preferences, 
shoppers are on the whole very similar in their overall statistical patterns, and 
return to stores with remarkable regularity: a Zipf's law $P(r) \sim r^{-\alpha}$ (with exponent $\alpha$ equal to 0.80 
and 1.13 for the North American and European datasets respectively) describes the 
frequency with which a customer visits a store at rank $r$ 
(where $r$ = 3 is his third most-frequented store, for example), independent 
of the total number of stores visited in a three-month period, see Figure~\ref{zipf}. This holds 
true despite cultural differences between the North American and Europe in consumption 
patterns and the use of credit cards. While our main focus is not the 
defense of any particular functional form or generative model of visitation 
patterns, our results support those of other studies showing the (power law) distribution 
of human and animal visitation to a set of sites~\cite{viswanathan1999optimizing,bartumeus2005animal,heisenberg2009free,doyle2009free}.

A universal measure of individual predictability would be useful in 
quantifying the relative regularity of a shopper. How much 
information is in a shopper's time series of consecutive stores? 

Informational entropy 
\cite{entropy} is commonly used to characterize the overall 
predictability of a system from which we have a time series of observations. It has also been 
used to show similarities and differences across individuals 
in a population~\cite{eagle}.

We consider two measures of entropy: \\

(i) The temporally-uncorrelated (TU) entropy for any individual $i$ 
is equal to $S_{TU}^{\alpha} = -\sum_{i \in M_{\alpha}} p_{\alpha,i} \log(p_{\alpha,i})$ 
where $p_{\alpha,i}$  is the 
probability that user $\alpha$ visited location $i$. Note this measure
is computed using only visitation frequencies, neglecting
the specific ordering of these visits. \\

(ii) The sequence-dependent (SD) entropy, which 
incorporates compressibility of the sequence of stores 
visited, is calculated using the Kolmogorov complexity estimate~\cite{lempel-ziv,li2008introduction}.

\begin{figure*}
\centering
\includegraphics[width=\textwidth]{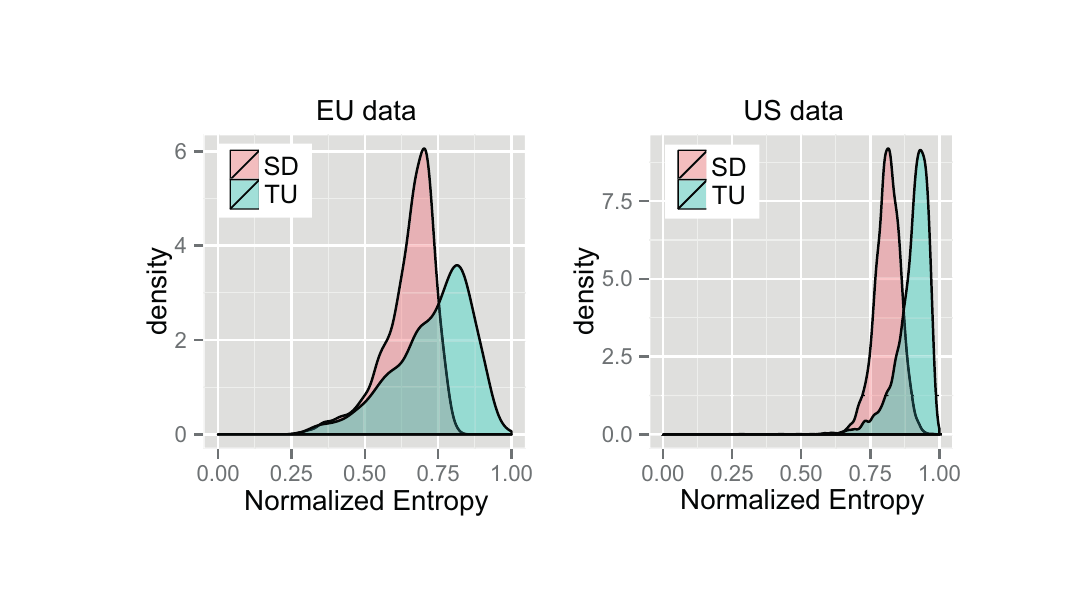}               
\caption{Normalized entropy distributions for the North American and European populations. Normalized entropy is computed by dividing the TU and SD entropies by the logarithm of the number of different merchants visited by a customer. TU entropy distributions are slightly higher for both populations}
\label{ents}
\end{figure*}

We find a narrow distribution of TU and SD entropies across each population, Figure \ref{ents}.

Another dataset, using cell phone traces \cite{song2010limits}, also finds a narrow distribution of entropies. This is not surprising, given the similarity of the two measures of individual trajectories across space. Yet we find a striking difference between the credit card and the cell phone data. In the shopping data, adding the sequence of stores (to obtain the SD entropy) has only a minor effect of the distribution, suggesting that individual choices are dynamic at the daily or weekly level. By contrast, cell phone data shows a larger difference. Why does this discrepancy occur? A possible 
explanation is that shoppers spread their visitation patterns more 
evenly across multiple locations than do callers. Even though visitation patterns
from callers and from shoppers follow a Zipf's law (figure 1), callers
are more likely to be found at a few most visited locations than are shoppers.
This is true, but to a point. Consumers visit their
single top location approximately $13\%$ (North American) and $22\%$ (European) of the time, while data from callers indicates more frequent visitation to top location.
 Yet shoppers' patterns  follow the same Zipf distribution seen in the cell phone data, and the narrow distribution of temporally-uncorrelated entropy
 indicates that shoppers are relatively homogenous in their behaviors.

An alternative explanation for our observed closeness of temporally-uncorrelated 
and sequence-dependent entropy distributions is the presence of small-scale 
interleaving and a dependence on temporal measurement. Over the 
course of a week a shopper might go first to the supermarket and 
then the post office, but he could just as well reverse this order. The ability to compare individuals is thus limited by the choice 
of an appropriate level of temporal resolution (not necessarily the same 
for each dataset) to sample the time series. With the large-scale mobility patterns inferred from cell phones, 
an individual is unable to change many routines: he drives to the 
office after dropping off the kids at school, while vice versa would
not be possible. In the more finite 
world of merchants and credit card swipes, there is space for 
routines to vary slightly over the course of a day or week.

To test the extent to which the second hypothesis explains the 
discrepancy between shoppers and callers, we simulate the effect 
of novel orderings by randomizing shopping sequence within a 24-hour period,
 for every day in our sample, and find little change in the 
measure of SD entropy. In other words, the re-ordering of 
shops on a daily basis does little to increase the predictability
 of shoppers, likely because the common instances of order 
swapping (e.g. coffee before rather than after lunch) are 
already represented in the data.  We then increase the sorting window from a single day to two days, to three days, and so forth.

Yet when we sort the order of shops visited over weekly intervals, 
thus imposing artificial regularity on shopping sequence, the true entropy is reduced significantly. 
If we order over a sufficiently long time period, we approach the 
values seen in mobile phone data. Thus entropy is a sampling-dependent 
measure which changes for an individual across time, depending on 
the chosen window. While consumers' patterns converge to very regular distributions
 over the long term, at the small scale shoppers are continually 
innovating by creating new paths between stores.

In order to measure the predictability of an individual's sequence of visits, we train
a set of first order Markov chain models. 
These models are based on the transition probabilities 
between different states, with the order of stores partially 
summarized in the first-order transition matrix. It is thus
related to the SD entropy measure.
We measure the probability of being at store $x$ at time $t+1$ as 
$Pr(X_{t+1} = x | X_{t} = x_t)$  and compare the prediction values 
to the observed values for each individual.
We build several models, varying the range of training data from 1 to 6 
months of data for each individual, and compare the model output 
to test data range of 1 to 4 subsequent months.

We additionally compare the results of the Markov models to the 
simplest naive model, in which the expectation is that an individual 
will chose his next store based on his distribution of visitation patterns, 
e.g. he will always go to one of the top two stores he visited most frequently in 
the training window (recall that for most people this store 
visitation frequency is on average just 20-35\%). Since this is a simply frequentist approach to the next-place prediction
problem, it is strongly related to TU entropy which is computed using the probability that a consumer visits a set of stores.

Comparing the match between model and observed data, we find that using
additional months of training does not produce significantly better
results. Moreover, results show some seasonal dependency (summertime and
December have lower prediction accuracy, for example). For fewer than three months 
of training data, the frequentist model does significantly better than the Markov model. This suggests the existence of a slow rate of  environmental change or exploration that would slowly undermine the model's accuracy.

For each of the two populations, we next test a global Markov model, in which all consumers' transition probabilities 
are aggregated to train 
the model. We find that such a model produces slightly better accuracy that either the naive or the individual-based 
models (with accuracy $\approx 25-27\%$). To test the 
sensitivity of this result we take ten global Markov models trained 
with $5\%$ of time series, selected randomly.  We find the standard deviation of the accuracy on these ten models increases to 3.6\% (from 0.3\% using all data), with similar mean accuracy. Thus the global Markov model depends on the sample of individuals chosen (for example, a city of connected individuals versus individuals chosen from 100 random small towns all over the world), but does in some cases add predictive power.

As previous work has indicated
\cite{deDomenico2012inter},
mobility patterns can be predicted with greater accuracy if we consider the traces of individuals with related behaviors.
In our case, even though we have 
no information about the social network of the customers, we can set a 
relationship between two people by analyzing the shared merchants they frequent. The global Markov model adds information about the plausible space of merchants that an
individual can reach, by analyzing the transitions of other customers that have
visited the same places, thus assigning a non-zero probability
to places that might next be visited by a customer.

Yet in almost every case, we find that people are in fact less predictable 
that a model based exclusively on their past behavior, or even that of their peers, would predict. 
In other words, people continue to innovate in the trajectories they 
elect between stores, above and beyond what a simple rate of 
new store exploration would predict.

\section{Discussion}

Colloquially, an unpredictable person can exhibit one of 
several patterns: he may be hard to pin down, reliably late, 
or merely spontaneous. As a more formal measure for human 
behavior, however, information-theoretic entropy conflates 
several of these notions. A person who discovers new shops 
and impulsively swipes his card presents a different case 
than the one who routinely distributes his purchases between 
his five favorite shops, yet both time series show a high TU entropy. 
Similarly, an estimate of the SD entropy can conflate a person 
who has high regularity at one level of resolution (for example, on a weekly basis) with one
 who is predictable at another. 

As example, take person A, who has the same schedule every
 week, going grocery shopping Monday evening and buying gas 
Friday morning. The only variation in A's routine is that he 
eats lunch at a different restaurant every day. On the other hand,
 person B sometimes buys groceries on Tuesdays, and sometimes 
on Sundays, and sometimes goes two weeks without a trip to 
the grocer. But every day, he goes to the local deli for lunch, 
after which he buys a coffee at the cafe next door. These 
individuals are predictable at different time-scales, but a 
global measure of entropy might confuse them as equally routine.

Entropy remains a useful metric for comparisons between 
individuals and datasets (such as in the present and cited studies), 
but further work is needed to tease out the correlates of 
predictability using measures aligned with observed behavior. 
Because of its dependence on sampling window and time intervals, 
we argue for moving beyond entropy as a measure of universal or even
of relative predictability. As our results suggest, models using entropy to measure predictability
are not appropriate for the small scale, that is, their individual 
patterns of consumption.

Shopping is the expression of both choice and necessity: we buy 
for fun and for fuel. The element of choice reduces an individual's predictability. In examining the solitary footprints that together comprise the
invisible hand, we find that shopping is a highly predictable
behavior at longer time scales. However, there exists substantial
unpredictability in the sequence of shopping events over short
and long time scales. We show that under certain conditions, even
perfect observation of an individual's transition probabilities
does no better than the simplistic assumption that he will go where
he goes most often.

\section{Methods}

\subsection{Data}

We sample tens of thousands individual accounts from one North American and one European financial institution. In the first case we represent purchases made by over 50 million accounts over a 6-month
window in 2010-2011; in the second, 4 million accounts in an 11-month window.
Data from transactions included timestamps with down-to-the-second resolution.

We filter each sample to best capture actual shoppers' accounts, to have sufficient data to train the Markov models with time series that span the entire time window, and to exclude corporate or infrequently used cards. We filter for time series in which the shopper 
visits at least 10 but no more than 50 unique stores in every month, and makes at least 50 but no more than 120 purchases per month. We test the robustness of this filter by comparing to a set of time series with an average of only one transaction per day (a much less restrictive filter), and find similar distributions of entropy for both filters.

The median and 25th/75th percentile merchants per customer in the filtered time series are  64, 46, and 87  in the North American (6 months) and  101, 69 and 131  in the European (11 months) dataset.

\subsection{Kolmogorov estimate of true entropy}
Kolmogorov entropy is a measure of the quantity of information needed
to compress a given time series by coding its component subchains. For instance, if a subchain appears several 
times within the series, it can be coded with the same symbol. The more repeated subchains exist, the less information is need to encode the series.

One of the most widely used methods to estimate 
Kolmogorov entropy is the Lempel-Ziv algorithm \cite{lempel-ziv,li2008introduction}, which measures SD entropy as: 
\begin{equation}
S_{SD}^{\alpha} \approx \frac{\log N}{\left< L(w) \right>}
\end{equation} 
where $\left< L(w) \right>$ is the average over the lengths of the 
encoded words. 

We can apply the algorithm to observed transitions between locations. A person with a smaller SD entropy 
is considered more predictable, as he is more constrained to the same sub-paths in the same order.

We can ignore the specific transition patterns between locations and simply define the
temporal-uncorrelated (TU) entropy as:
\begin{equation}
S_{TU}^{\alpha} = -\sum_{i \in M_{\alpha}} p_{\alpha,i} \log(p_{\alpha,i})
\end{equation}
Note that random messages over a set of sub-cahins should satisfy 
$S_{SD}^{\alpha} \approx S_{TU}^{\alpha}$ with a large enough 
length of message. Given the distributions of the temporal-uncorrelated and 
Kolmogorov entropies across a population, we can analyze the predictability 
of the mobility patterns by statistically comparing these two measures.

\subsection{Markov  model}
Markov chains are used to model temporal stochastic processes, in which the present state depends only on the previous one(s). Mathematically, let $X_t$ be a sequence
of random variables such that
\begin{align}
 \nonumber&P(X_t = x_t | X_{t-1} = x_{t-1}, X_{t-2} = x_{t-2},...) = \\
 &P(X_t = x_t | X_{t-1})
\end{align}
then $\left\lbrace X_t \right\rbrace$ is said to be a Markov process of first order.
This process is summarized with transition matrix $P=(p_{ij})$ 
where $p_{ij} = P(X_t=x_j |X_{t-1}=X_i)$.
 Markov chains can be considered an extension of a simple frequentist model in which
\begin{align}
 \nonumber&P(X_t = x_t | X_{t-1} = x_{t-1}, X_{t-2} = x_{t-2},...) = \\
 &P(X_t = x_t )
\end{align}
applied on every observed state. 

If the present transaction location depends in some part on the previous one, a 1st order Markov model would be able to predict the location with greater accuracy than a simple frequentist model.

Our methods allow us to note two relationships:
\begin{itemize}
 \item \emph{Temporal-uncorrelated entropy and frequentist model:} both use $P(X_t = x_t)$ without additional information. Temporal-uncorrelated is a good approximation of the distribution of states, and is thus related to the performance of the
  frequentist model.
 \item \emph{Sequence-dependent entropy and Markov model:} SD entropy is a single measure of all sub-chain frequencies, and is thus related to the accuracy of a 1st order Markov model, which represents the probability of a single set of sub-chains occuring.
\end{itemize}

\section{Author Contributions}

Developed the ideas, methods, and analysis: CK, ALP, MC, SP, EME. Analyzed the data: CK, ALP. Wrote the manuscript: CK, ALP, MC, EME

\section{Competing Financial Interests}

The authors declare they have no competing financial interests.



\newpage

\begin{figure*}
 \centering
 \includegraphics[width=\textwidth]{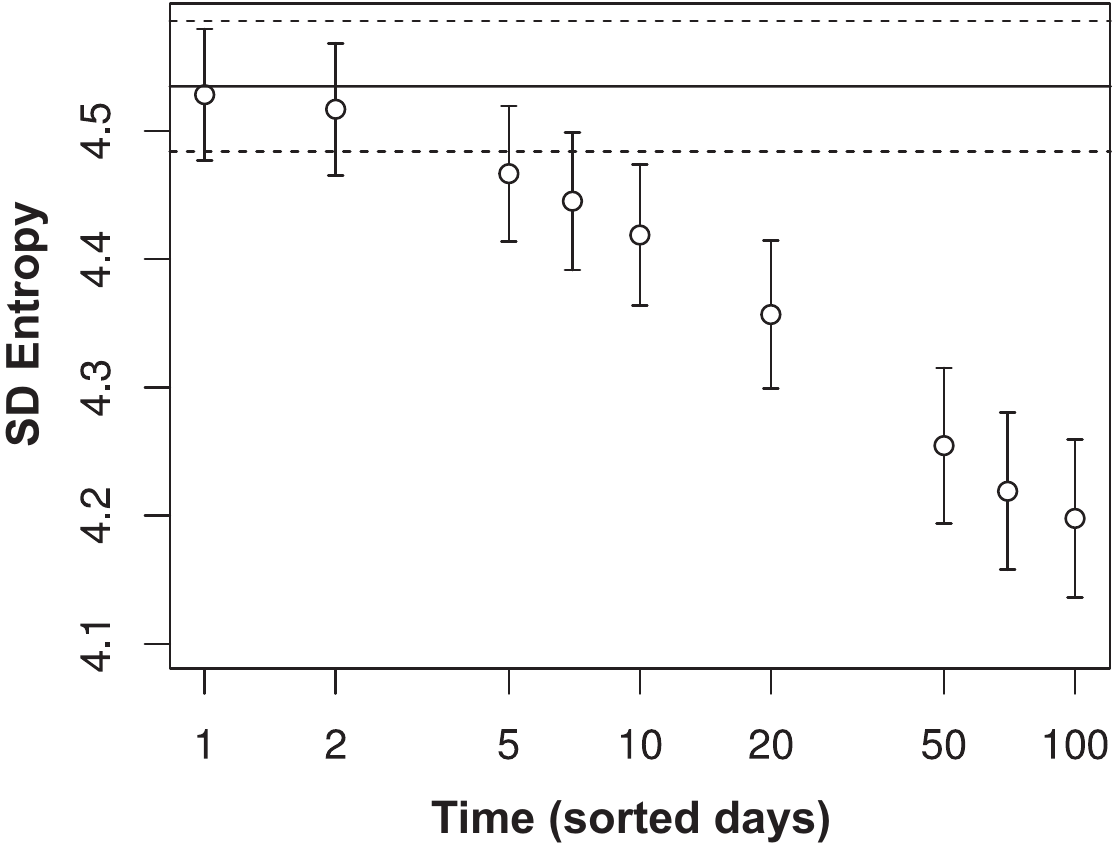}
 \caption{sequence-dependent entropy for a number of "artificially sorted" sequences. For each window size over which the time series is sorted, we measure the sequence-dependent entropy for the population and estimate the error of the mean. 
 The horizontal solid line at the top of the figure
 indicates the average SD entropy for the original data whereas the dashed lines depict the band for the error of the mean.}
 \label{sorted_ent}
\end{figure*}

\begin{figure*}
\centering
 \includegraphics[totalheight=1.0\textwidth, angle=-90]{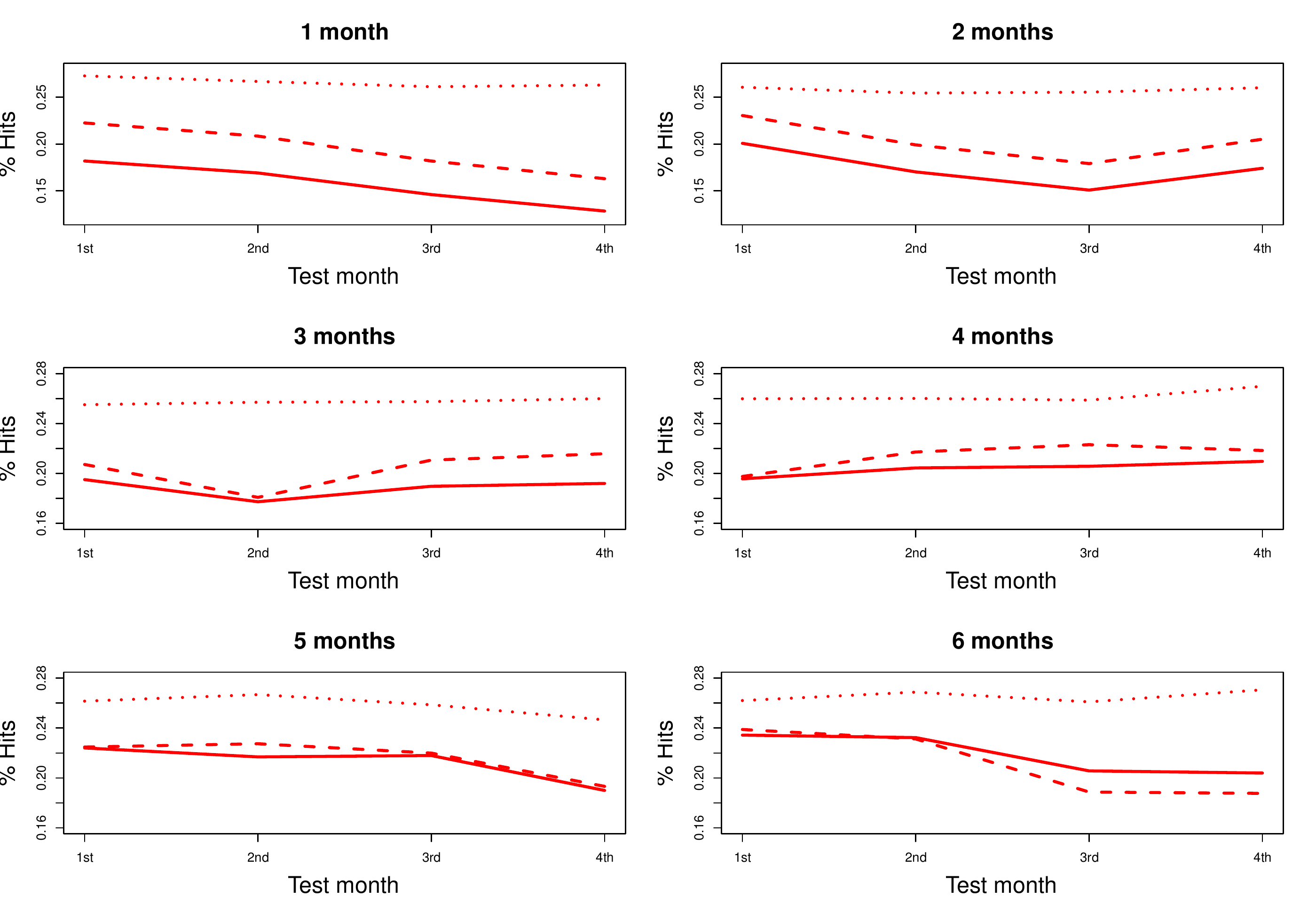}
  \caption{Markov model results for different temporal windows in training and test.
  The solid red line indicates hit percentage for Markov model, dashed line exhibits
  accuracy for the naive model and the pointed line indicates results for the 
  Global Markov model.}
  \label{markov1}
\end{figure*}

\end{document}